\documentclass[
    ,final            
  ,numberedheadings 
  ]
  {aipproc}

\layoutstyle{8x11double}

\begin{document}

\title{Two-point functions of quenched lattice QCD \\
in Numerical Stochastic Perturbation Theory}



\author{F. Di Renzo}{
  address={Dipartimento di Fisica, Universit\`a di Parma and INFN, I-43100 Parma, Italy}
}

\author{E.-M. Ilgenfritz}{
  address={Institut f\"ur Physik, Humboldt-Universit\"at zu Berlin, D-12489 Berlin, Germany}
}

\author{H. Perlt}{
  address={Institut f\"ur Theoretische Physik, Universit\"at Leipzig, D-04009 Leipzig, Germany}
}

\author{{A. Schiller\footnote{Speaker}~}}{
  address={Institut f\"ur Theoretische Physik, Universit\"at Leipzig, D-04009 Leipzig, Germany}
}

\author{C. Torrero}{
  address={Institut f\"ur Theoretische Physik, Universit\"at Regensburg, D-93040 Regensburg,  Germany}
}

\begin{abstract}
We summarize the higher-loop  perturbative computation
of the ghost and gluon propagators in $SU(3)$ Lattice Gauge Theory. 
Our final aim is to compare with results from lattice simulations in 
order to expose the genuinely non-perturbative content of the latter. 
By means of Numerical Stochastic Perturbation Theory we compute the ghost 
and gluon propagators in Landau gauge up to three and four loops.
We present results in the infinite volume and $a \to 0$ limits, based on 
a general fitting strategy.
\end{abstract}

\maketitle

\section{Introduction}
\vspace{-2mm}

This talk  summarizes our work on the 
higher-loop perturbative gluon and ghost propagators in Landau 
gauge~\cite{DiRenzo:2009ni,DiRenzo:2010cs}. The Monte Carlo study of both 
propagators, which are closely related to each other by Schwinger-Dyson 
equations (SDE), has attracted much attention outside the lattice community 
by phenomenologists working 
on infrared QCD in general and hadron physics (see our original papers for 
further references). 
Taken together, both propagators provide us with a definition
and the momentum dependence of the running coupling $\alpha_s(q^2)$ 
directly based on the ghost-gluon vertex. 

A simple connection between the two propagators exists in the extreme infrared, 
both beeing powerlike in a scaling or massive in a decoupling solution. 
This nonuniqueness reflects the Gribov problem.
The effect of nontrivial vacuum structure (vortices, instantons) is manifest
also in the gluon propagator, in the intermediate momentum range around 
${\cal O}(1 \mathrm{~GeV})$ where the SDE approach suffers from truncation 
ambiguities and where nonperturbative lattice calculations are unrivalled.
In order to follow the onset of nonperturbative effects, it is desirable
to approach this momentum range from high momenta within higher-order 
perturbation theory. While ordinary diagrammatic lattice perturbation theory 
(LPT) soon gets too involved to be pursued, Numerical Stochastic Perturbation Theory 
(NSPT, for a recent review see Ref.~\cite{DiRenzo:2004ge} and references therein), 
provides a powerful tool to perform high-loop computations.

\section{NSPT in a nutshell}
\vspace{-2mm}

NSPT has its roots in stochastic quantization and is based
on a modified Langevin equation equipped with stochastic gauge fixing.
We use here a version for quenched lattice QCD with Wilson gauge action.
Actually, it is a hierarchy of first-order evolution equations associated with 
various parts of the gauge link fields $U$ and gluon fields $A$ exposed by an
expansion 
in powers of the lattice coupling $g \propto 1/\sqrt{\beta}$:
\begin{eqnarray}
 U_{x,\mu} = \sum\limits_{l \ge 0} \beta^{-l/2} U^{(l)}_{x,\mu} \,, \ \
 A_{x+\frac{\hat{\mu}}{2},\mu} =
    \sum\limits_{l \ge 1} \beta^{-l/2} A^{(l)}_{x+\frac{\hat{\mu}}{2},\mu} \,. 
\end{eqnarray}
These different orders are separately dealt within the code. The maximal 
addressable order of perturbation theory is thus limited by the available 
computing resources (cpu time and memory).

The Langevin simulation is implemented in an Euler scheme with a finite evolution 
time step. Before the 
estimator for the gauge dependent ghost and gluon propagators
can be evaluated, we have to fix the gauge to the minimal Landau gauge.
For this purpose, a sequence of configurations (separated by ${\cal O}(50)$
Langevin time steps) 
is subjected to a Fourier-accelerated gauge-fixing procedure, after which the 
individual gluon fields, $A^{(l)}_\mu$ (associated with particular perturbative 
order $g^l$) are transversal within machine precision.

The propagators are evaluated taking the long-time average of coefficients, 
order by order in a loop expansion in even powers of $g$. Contributions from odd
powers vanish within the statistical errors. As for any Langevin simulation, one 
then has to take the limit to vanishing time step. 
In order to get results comparable with the practice of LPT, the continuum limit 
and the limit of infinite volume must be performed.
NSPT results for {\it finite} lattice volume and spacing can be confronted 
directly with standard MC results for a given $\beta$, 
provided the definitions of the studied observables is the same.

The gluon two-point function in $n$-loop order is defined as a
convolution of the bilinears of gluon fields (in momentum space) in
complementary orders ($A_\mu ^{(l)}=T^a A^{a,(l)}$, $p_\mu (k_\mu)=2 \pi k_\mu/(a N)$):
\begin{equation}
  \delta^{ab} D_{\mu\nu}^{(n)}(p(k)) = \left\langle \,
  \sum_{l=1}^{2n+1}
  \left[ \widetilde{A}^{a,(l)}_{\mu}(k) \,
  \widetilde{A}^{b,(2n+2-l)}_{\nu}(-k) \right] \,
  \right\rangle_U \,.
\label{eq:Dn}
\end{equation}
In Landau gauge we consider
$\sum_{\mu=1}^4 D_{\mu\mu}^{(n)}\equiv 3 D^{(n)}$
and use the dressing functions 
($ \hat p_{\mu}(k_{\mu}) = ({2}/{a}) \sin\left({\pi k_{\mu}}/{N}\right)$)
\begin{equation}
  J^{G,(n)}(p) = p^2 \; D^{(n)}(p(k)) \,, 
  \hat J^{G,(n)}(p) = \hat p^2 \; D^{(n)}(p(k)) \,.
\end{equation}

The color diagonal ghost propagator in momentum space is 
the color trace in the adjoint representation
\begin{equation}
  G( p(k)  )=
  \frac{1}{8} \left\langle {\rm{Tr_{adj}}}~M^{-1}(k)\right\rangle_U \,.
  \label{eq:G-definition}
\end{equation}
In (\ref{eq:G-definition}) $M^{-1}(k)$ is the Fourier transform of the inverse
FP operator in lattice coordinate space.
It is expanded in terms of products of various $A^{(l)}$,
with the term $M^{(j)}$ collecting all terms of order $g^j$.
This structure allows to express $M^{-1}(k)$ also as an expansion in orders 
of $g$ in a recursive way.
Again we use the ghost dressing functions
\begin{equation}
  J^{(n)}(p) = p^2 \; G^{(n)}(p(k)) \,, 
  \hat J^{(n)}(p) = \hat p^2 \; D^{(n)}(p(k)) \,.
\end{equation}

\begin{figure}[!htb]
  \centering
  \includegraphics[scale=0.5,clip=true]{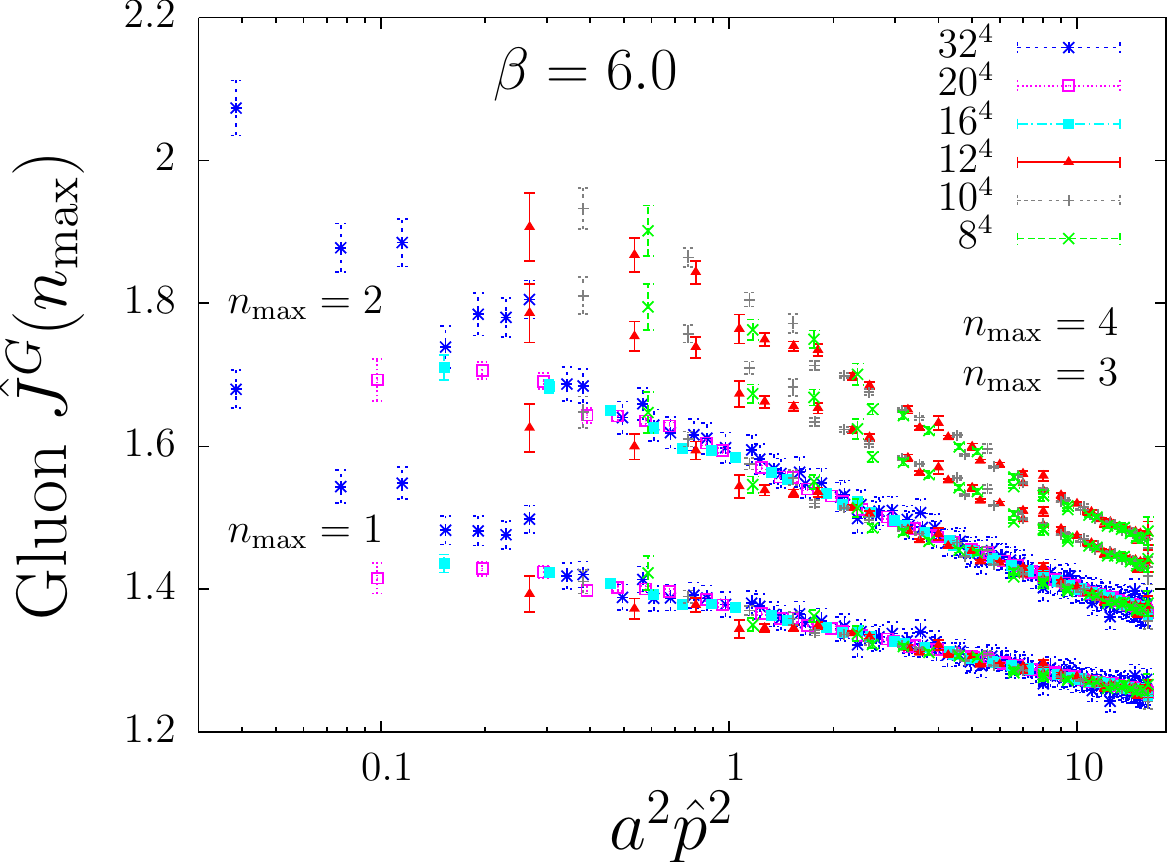}
  \vspace{-8mm}
  \caption{The cumulatively summed perturbative gluon dressing function for various volumes.}           
  \label{fig:figure15b}
\end{figure}

As an example the cumulatively summed perturbative gluon dressing function 
for various volumes
is shown in Fig.~\ref{fig:figure15b} using $g^2=6/\beta=1$.

A reasonable ``convergence'' of the NSPT results up to few loops
(three or four are available now) requires a small bare coupling $g$.
However, $g$ is known to be
a poor expansion parameter~\cite{Lepage:1992xa}. One can speed up
convergence by ``boosting'', i.e., trading the bare coupling constant
by an effective ``boosted'' coupling $g_b^2 =
g^2/P_{\mathrm{pert}}(g^2) > g^2$. 
Here $P_{\mathrm{pert}}$ is defined by the average perturbative plaquette
determined also within our Langevin simulations.
The effect of the boosted coupling being larger is overcompensated by the 
rapid decay of the expansion coefficients with increasing order $n$.

\begin{figure}[!htb]
  \centering
   \includegraphics[scale=0.5,clip=true]{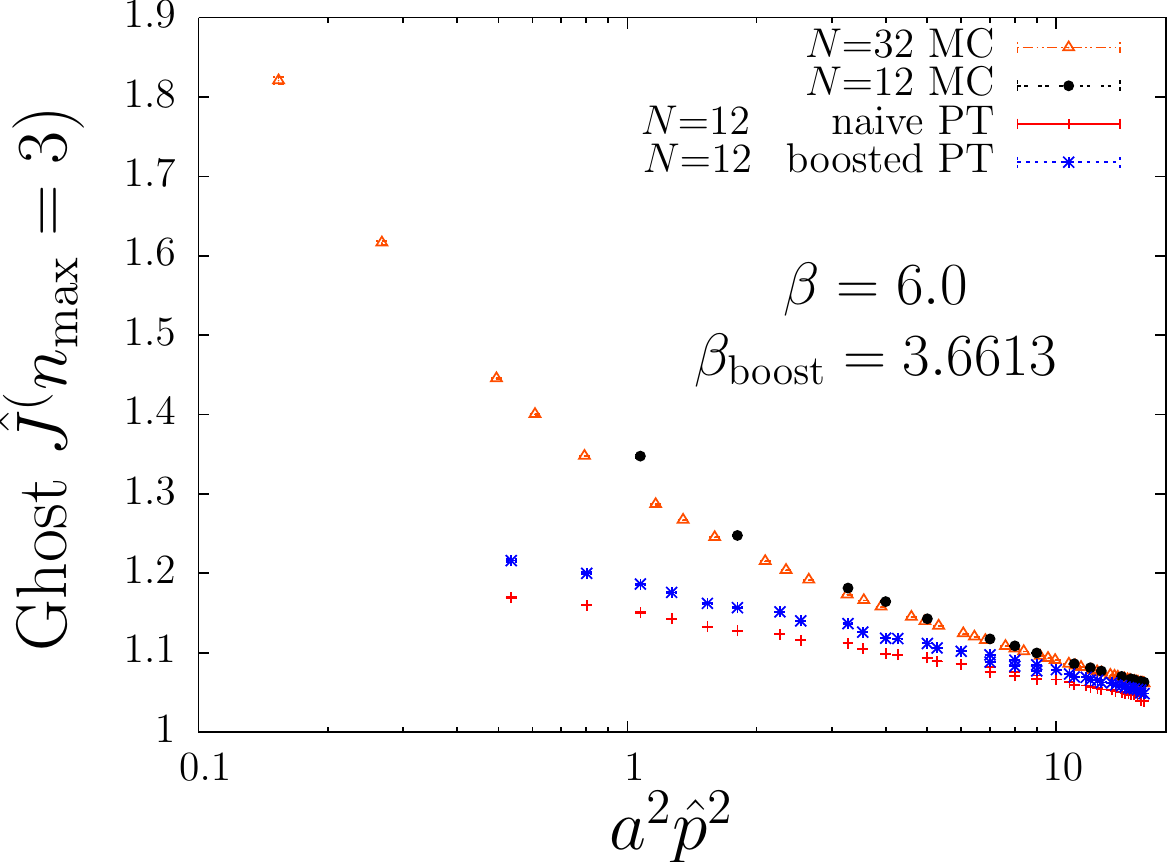}
  \vspace{-8mm}
   \caption{
           The ghost dressing function in three loops for naive and boosted
           NSPT comapred to MC data for a $12^4$ lattice.
           }
   \label{fig:figure16a}
\end{figure}

\begin{figure}[!htb]
  \centering
    \includegraphics[scale=0.5,clip=true]{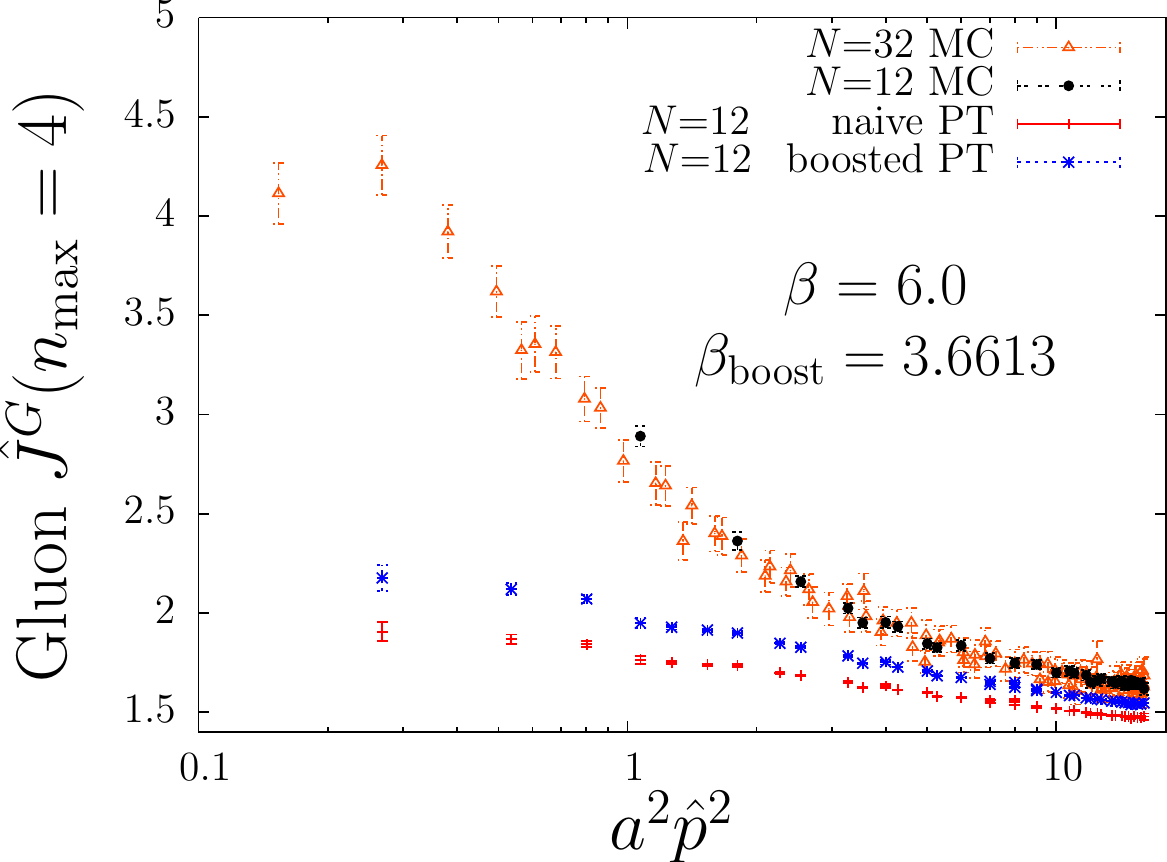}
  \vspace{-8mm}
  \caption{
           Same as in  Fig.~\ref{fig:figure16a} for the gluon dressing function in four loops.
          }
   \label{fig:figure16b}
\end{figure}

We illustrate the effect of ``boosting'' the perturbative expansion and confront 
the boosted dressing functions with corresponding new Monte Carlo (MC) data of the 
Berlin group~\cite{Ilgenfritz:2010gu} adopting 
the same definitions for the propagators and the gauge fixing as in NSPT. This 
is shown in Figs.~\ref{fig:figure16a},\ref{fig:figure16b} where also the bare and the
boosted
inverse couplings $\beta$ and $\beta_{\mathrm{boost}}$ 
are given. As expected, boosting moves the NSPT data closer to the MC results, 
but they cannot be reached completely, certainly not at $\beta=6.0$.
\vspace{-5mm}
  \vspace{-5mm}

\begin{figure}[!htb]
  \centering
   \includegraphics[scale=0.5,clip=true]{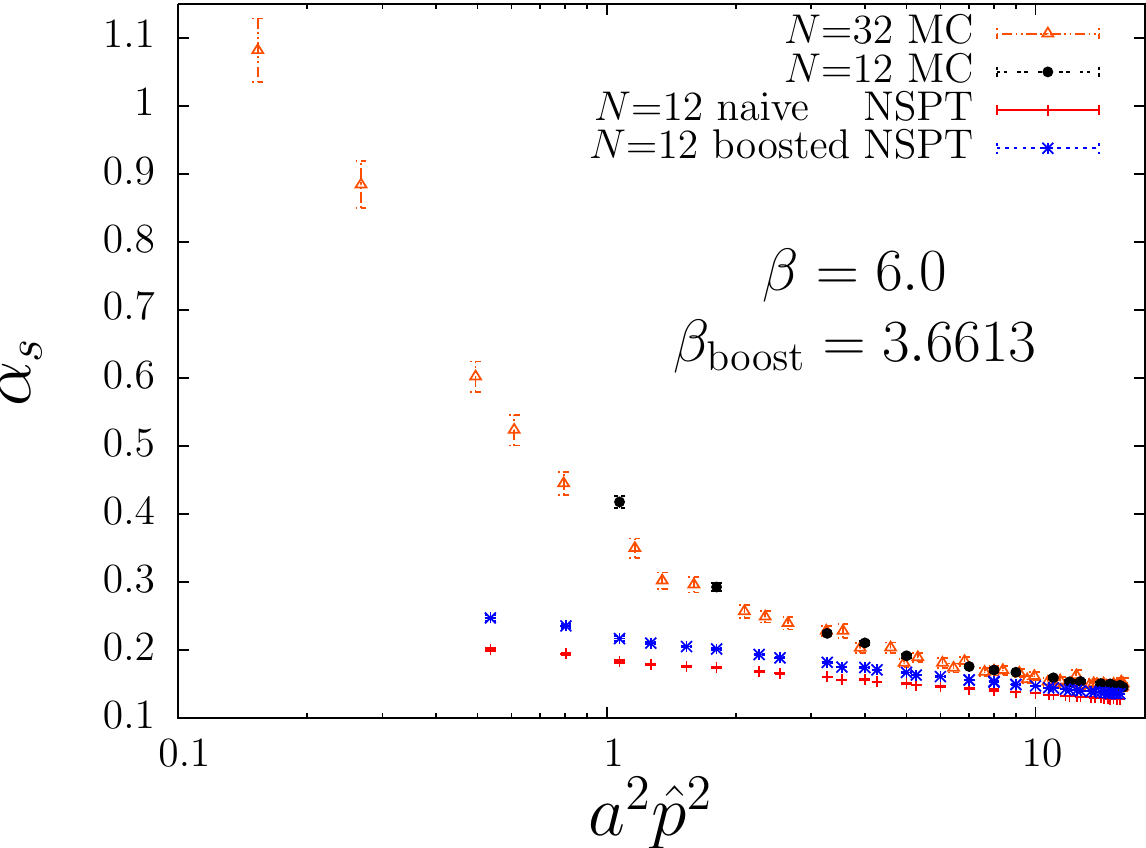}
   \caption{
           Comparing naive and boosted PT (based on NSPT data)
           for the running coupling constant $\alpha_s$
           to corresponding MC data for a $12^4$ lattice at $\beta=6.0$. 
           }
   \label{fig:figure18a}
\end{figure}

\begin{figure}[!htb]
  \centering
    \includegraphics[scale=0.5,clip=true]{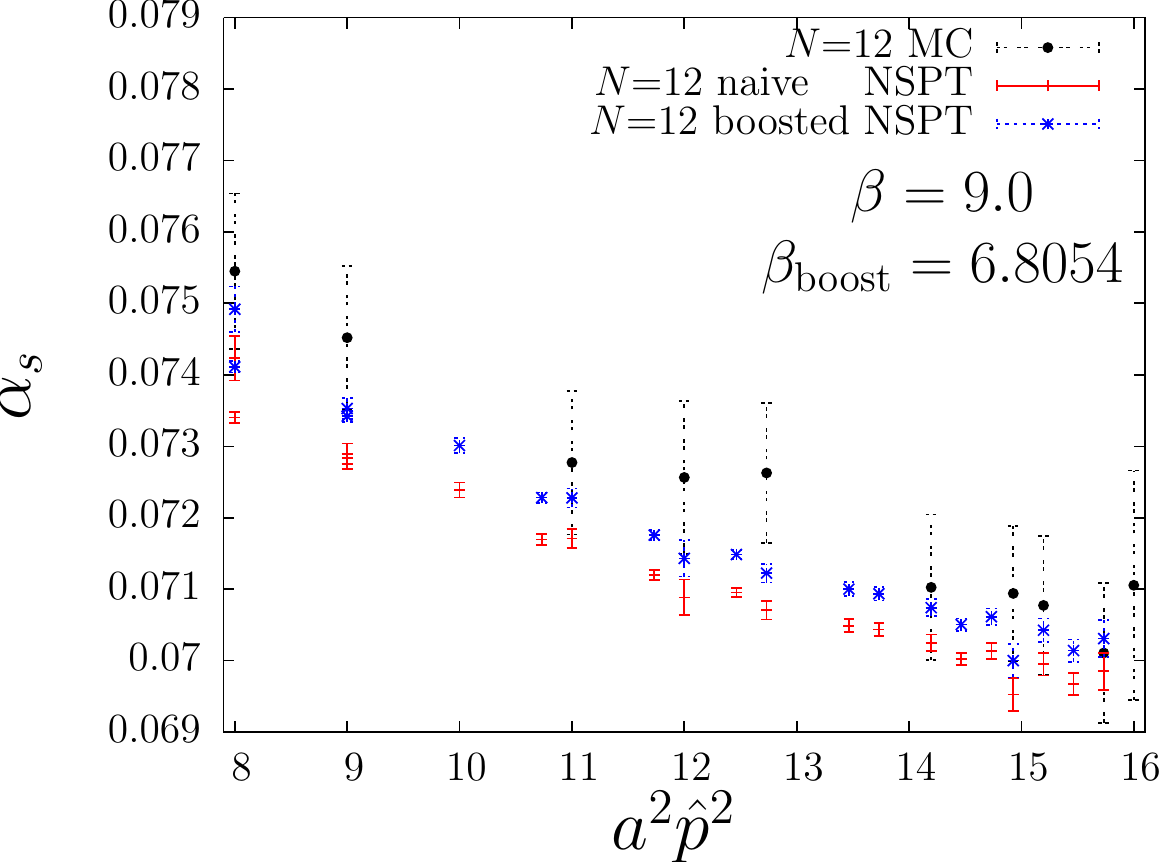}
  \caption{
           Same as in  Fig.~\ref{fig:figure18a} at $\beta=6.0$ zoomed into the large momentum region.
          }
   \label{fig:figure18b}
\end{figure}

Here we define the renormalization-group invariant running coupling $\alpha_s$ 
by the ghost-gluon vertex in a particular (minimal) MOM scheme 
(see e.g.~\cite{Alkofer:2000wg}). It is given  in terms of the bare gluon and ghost dressing
functions $J^{\rm G}$ and $J$ as follows:
\begin{equation}
 \alpha_s (p(k)) = \frac{6}{4 \pi \beta}\,  \hat J(p(k),\beta)^2 \,  \hat J^G(p(k),\beta) \,.
  \label{eq:running_coupling}
\end{equation}
The $\alpha_s$ calculated from the NSPT dressing functions, both summed up to the
orders available, is compared to the MC results at
$\beta=6.0$ and $9.0$. The corresponding data is shown in
Figs.~\ref{fig:figure18a},\ref{fig:figure18b} 
again for naive and boosted perturbation theory.
We see that the running coupling from MC simulations
is approached at large momenta from below up to 7\% for $\beta=6.0$ and practically
approached within the present errors for $\beta=9.0$, i.e. in an
effectively deconfined phase.
\vspace{-4mm}

\section{Dressing functions in the limits $V\to \infty$ and $ap \to 0$}
\vspace{-2mm}

In the RI'-MOM scheme the dressing functions are just the 
wave function renormalization constants for the ghost and gluon fields at the 
renormalization point $\mu^2=p^2$.
The NSPT data available at various volumes and lattice momentum realizations
allow us to find the perturbative dressing functions  
to three-loop accuracy in the bare coupling including the non-logarithmic 
contributions. Via standard transformations the results can be transformed to 
the renormalized coupling in the 
preferred scheme. To find those constants, a fitting procedure has been proposed 
which takes into account both hypercubic and finite volume effects (for details 
see ~\cite{DiRenzo:2009ni}).
As result we get ($L \equiv \log (pa)^2$)
\begin{eqnarray}
\label{J3loopNum}
&&J(a,p,\beta) = 1 + \frac{1}{\beta}\Bigl[ -0.0854897\, L + 0.525314 \Bigr]
  \nonumber \\
& & 
 + \frac{1}{\beta^2}\Bigl[ 0.0215195\, L^2  - 0.358423 \, L   + 1.4872(57) \Bigr] 
\nonumber \\
& & 
 + \frac{1}{\beta^3}\Bigl[ -0.0066027 L^3  + 0.175434\, L^2 
 \\
& & \hspace{-0.1cm}
-1.6731(1)  \, L + 4.94(27) \Bigr]
\nonumber
\end{eqnarray}
\begin{eqnarray}
  \label{J3GloopNum}
  &&J^{G}(a,p, \beta) = 1 + \frac{1}{\beta}\Bigl[ -0.24697\, L  + 2.29368 \Bigr]
   \nonumber \\
  & & 
  + \frac{1}{\beta^2}\Bigl[ 0.08211 L^2 - 1.48445  \, L  + 7.93(12) \Bigr] 
  \nonumber \\
  & &  
  + \frac{1}{\beta^3}\Bigl[ -0.02964 L^3  + 0.81689 \, L^2 
 \\
& & \hspace{-0.1cm}
 - 8.13(3) \, L + 31.7(5) \Bigr]
  \nonumber
\end{eqnarray}
The results for one-loop lattice perturbation theory are known for a long time,
the higher-loop non-leading log's and constant contributions are our predictions
for the Landau gauge. 
\vspace{-4mm}

\section{Conclusion}
\vspace{-2mm}

We have calculated the gluon propagator in Landau gauge up to four 
and the ghost propagator up to three loops in NSPT.
The dressing functions summed using boosted PT 
are compared to recent MC measurements of the
Berlin Humboldt University group. 
At large lattice momenta the dressing functions
with more than four loops will match the MC measurements, thus enabling a fair
accounting of the perturbative tail 
taking care of discretization effects. This can be used as an 
alternative to fitting the high momentum tail of MC results by continuum-like 
formulae.

We worked out the relation to standard LPT in limits $V\to \infty$ and $pa \to 0$.
For this aim we developed a fitting strategy for lattice artifacts and 
finite-size corrections.  
We find good agreement with known one-loop results of diagrammatic LPT
and present original two- and three-loop results for the propagators.


\vspace{-3mm}

\begin{thebibliography}{6}
\expandafter\ifx\csname natexlab\endcsname\relax\def\natexlab#1{#1}\fi
\providecommand{\enquote}[1]{``#1''}
\expandafter\ifx\csname url\endcsname\relax
  \def\url#1{\texttt{#1}}\fi
\expandafter\ifx\csname urlprefix\endcsname\relax\def\urlprefix{URL }\fi
\providecommand{\eprint}[2][]{\url{#2}}

\bibitem[Di~Renzo et~al.(2010)]{DiRenzo:2009ni}
F.~Di~Renzo, E.-M. Ilgenfritz, H.~Perlt, A.~Schiller, and C.~Torrero,
  \emph{Nucl. Phys.} \textbf{B831}, 262--284 (2010), \eprint{0912.4152}.

\bibitem[Di~Renzo et~al.(2011)]{DiRenzo:2010cs}
F.~Di~Renzo, E.-M. Ilgenfritz, H.~Perlt, A.~Schiller, and C.~Torrero,
  \emph{Nucl. Phys.} \textbf{B842}, 122--139 (2011), \eprint{1008.2617}.

\bibitem[Di~Renzo and Scorzato(2004)]{DiRenzo:2004ge}
F.~Di~Renzo, and L.~Scorzato, \emph{JHEP} \textbf{10}, 073 (2004),
  \eprint{hep-lat/0410010}.

\bibitem[Lepage and Mackenzie(1993)]{Lepage:1992xa}
G.~P. Lepage, and P.~B. Mackenzie, \emph{Phys. Rev.} \textbf{D48}, 2250--2264
  (1993), \eprint{hep-lat/9209022}.

\bibitem[Ilgenfritz et~al.(2010)]{Ilgenfritz:2010gu}
E.-M. Ilgenfritz, C.~Menz, M.~M\"uller-Preussker, A.~Schiller, and A.~Sternbeck
  (2010), \eprint{1010.5120}.

\bibitem[Alkofer and von Smekal(2001)]{Alkofer:2000wg}
R.~Alkofer, and L.~von Smekal, \emph{Phys. Rept.} \textbf{353}, 281--465
  (2001), \eprint{hep-ph/0007355}.

\end{thebibliography}
\end{document}